\documentclass[superscriptaddress,twocolumn,showpacs,aps,nofootinbib,prl]{revtex4}

\usepackage{amssymb,xcolor}
\usepackage{amsmath}
\usepackage{amsbsy}
\usepackage{amsthm}
\usepackage{graphicx}
\usepackage{amsfonts}
\usepackage{epstopdf}
\usepackage{times}
\usepackage{txfonts}
\usepackage{hyperref}

\newcommand{\beq}{\begin{equation}}
\newcommand{\eeq}{\end{equation}}

\newcommand{\bra}[1]{\langle#1|}

\newcommand{\ket}[1]{|#1\rangle}

\newcommand{\ketbra}[3]{|{#1}\rangle_{#2}\langle{#3}|}
\newcommand{\id}{\leavevmode\hbox{\small1\normalsize\kern-.33em1}}

\newcommand{\clock}{{\circlearrowright}}
\newcommand{\cclock}{{\circlearrowleft}}

\begin{document}

\title{Experimental Quantum Networking Protocols via Four-Qubit Hyperentangled Dicke States}

\author{A. Chiuri}
\affiliation{Dipartimento di Fisica, Sapienza {Universit\`a} di Roma, Piazzale Aldo Moro 5, I-00185 Roma, Italy}
\author{C. Greganti}
\affiliation{Dipartimento di Fisica, Sapienza {Universit\`a} di Roma, Piazzale Aldo Moro 5, I-00185 Roma, Italy}
\author{M. Paternostro}
\affiliation{Centre for Theoretical Atomic, Molecular, and Optical Physics, School of Mathematics and Physics, Queen's University, Belfast BT7 1NN, United Kingdom}
\author{G. Vallone}
\affiliation{Department of Information Engineering, University of Padova, I-35131 Padova, Italy}
\author{P. Mataloni}
\affiliation{Dipartimento di Fisica, Sapienza {Universit\`a} di Roma, Piazzale Aldo Moro 5, I-00185 Roma, Italy}
\affiliation{Istituto Nazionale di Ottica (INO-CNR), L.go E. Fermi 6, I-50125 Firenze, Italy}

\date{\today}

\begin{abstract}
We report the experimental demonstration of two quantum networking protocols, namely quantum $1{\rightarrow}3$ telecloning and open-destination teleportation, implemented using a four-qubit register whose 
state is encoded in a high-quality two-photon hyperentangled Dicke state. The state resource is characterized using criteria based on multipartite entanglement witnesses. We explore the characteristic 
entanglement-sharing structure of a Dicke state by implementing high-fidelity projections of the four-qubit resource onto lower-dimensional states.  Our work demonstrates for the first time the usefulness 
of Dicke states for quantum information processing. 
 
\end{abstract}

\pacs{
42.50.Dv,
03.67.Bg,
42.50.Ex
}

\maketitle

Networking offers the benefits of connectivity and sharing, often allowing for tasks that individuals are unable to accomplish on their own. This is known for computing, where grids of processors outperform the computational power of single machines or allow the storage of much larger databases. It should thus be expected that similar advantages are transferred to the realm of quantum information. 
Quantum networking, where a given task is pursued by a lattice of local nodes sharing (possibly entangled) quantum channels, is emerging as a realistic scenario for the implementation of quantum protocols requiring medium/large registers. Key examples
of such approach are given by quantum repeaters~\cite{repeaters}, non-local gates~\cite{distributed}, scheme for light-mediated interactions of distant matter qubits~\cite{kimble} and one-way quantum computation~\cite{Briegelreview}. 

In this scenario, photonics is playing an important role: the high reconfigurability of photonic setups and outstanding technical improvements have facilitated the birth of a new generation of experiments (performed both in bulk optics and, 
recently, in integrated photonic circuits~\cite{integrato}) that have demonstrated multi-photon quantum control towards high-fidelity computing with registers of a size inaccessible until only recently~\cite{vall08prl,lu07nap,gao10prl,vall10pra3,bigge09prl,gao10nap}.
The design of complex interferometers and the exploitation of multiple degrees of freedom of a single photonic information carrier have enabled the production of interesting states, such as cluster/graph states, GHZ-like states and (phased) 
Dicke states~\cite{kies07prl,dickeexp,dickeexpRome}, among others~\cite{bourennane2010,bour06prl}. Dicke states have been successfully used to characterize multipartite entanglement close to fully symmetric states and its robustness to decoherence~\cite{dickeexpRome}. They  are potentially useful resource for the implementation of protocols for distributed quantum 
communication such as quantum secret sharing~\cite{hillery}, quantum telecloning (QTC)~\cite{mura99pra}, and open destination teleportation (ODT)~\cite{bourennane,panODT}. So far, such opportunities have only been examined theoretically and confirmed indirectly~\cite{kies07prl,dickeexp}, 
leaving a full implementation of such protocols unaddressed.
 
\begin{figure}
\includegraphics[width=\linewidth]{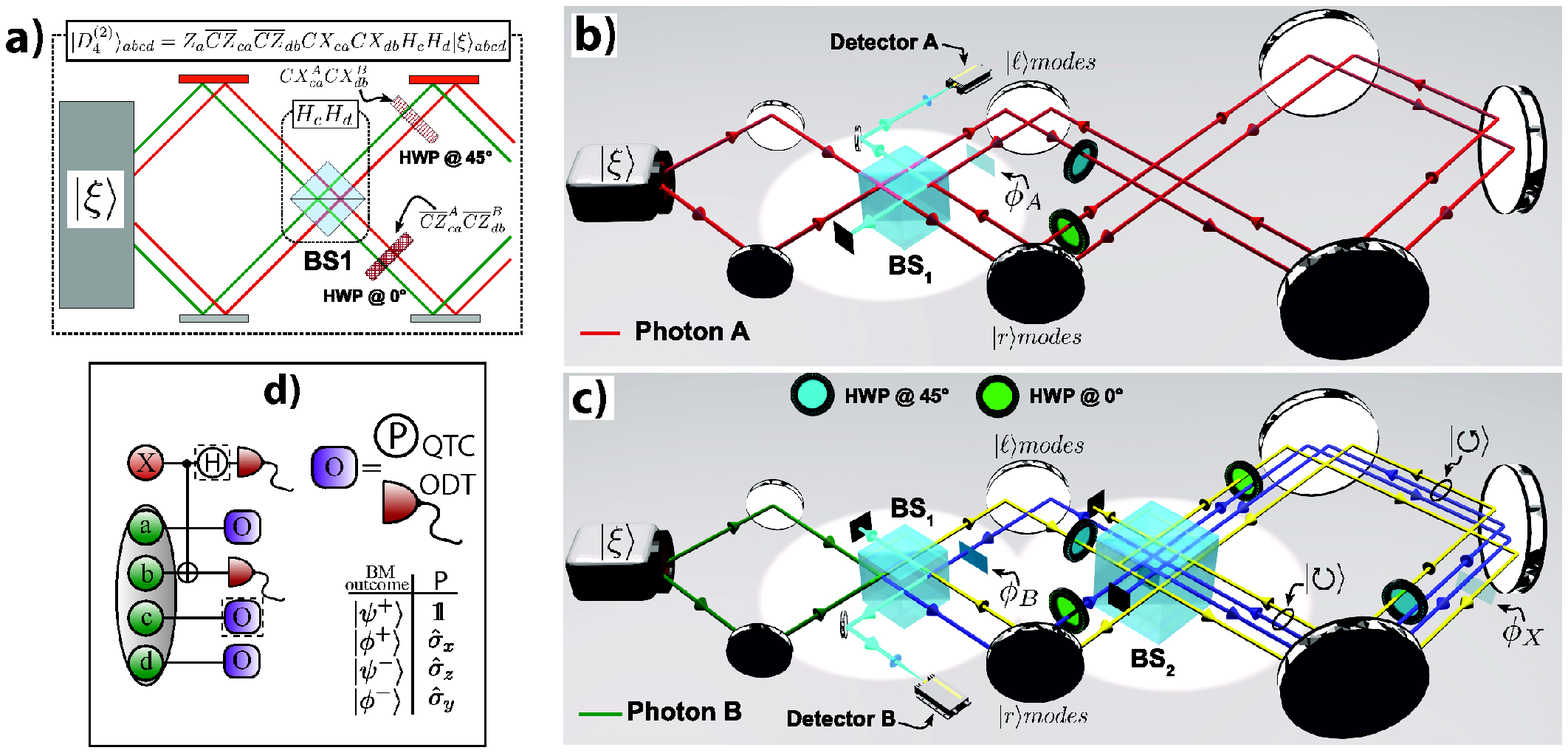}
\caption{
{\bf a)} Scheme for the $\ket{\xi}\to\ket{D^{(2)}_4}$ conversion. The spatial qubits experience
the Hadamard gates ${\sf H}_{c,d}$ implemented through a polarization insensitive beam splitter (${\rm BS}_1$). A controlled-NOT (controlled-PHASE) gate
${\sf CX}{=}\ketbra{0}{i}{0}\otimes\openone_{j}+\ketbra{1}{i}{1}\otimes\hat{\sigma}^x_j$ ($\overline{\sf CZ}{=}\ketbra{1}{i}{1}\otimes\openone_{j}+\ketbra{0}{i}{0}\otimes\hat{\sigma}^z_j$) is realized  by a half-wave plate (HWP) with axis at $45^\circ$ ($0^\circ$) with respect to the vertical direction~$(i{=c,d},~j{=}a,b$). The control (target) qubit of such gate is the path (polarization) degree of freedom (DOF). {\bf b)} \& {\bf c)} Displaced Sagnac loop for the realization 
of the QTC/ODT protocol. Panel {\bf b)} [{\bf c)}] shows the path followed by the upper [lower] photon A [B]. The glass plates $\phi_{A,B,X}$  allow us to vary the relative phase between the different paths within the interferometer. {\bf d)} Circuit for $1{\rightarrow}3$ QTC and ODT. 
Qubits $\{a,b,c,d\}$ are prepared in $\ket{D^{(2)}_4}$ while $X$ should be cloned/teleported. 
For QTC, the ${\sf CX}_{Xb}$ gate is complemented by the projection of $X$ ($b$) on the eigenstates of $\sigma^x$ ($\sigma^z$), so as to perform a BM. For QTC (ODT), operation O is a local Pauli gate ${\sf P}$ determined by the outcome of the BM according 
to the given table. 
For ODT (with, say, receiver qubit $c$), the operations in the dashed boxes should be removed.}
\label{setup}
\end{figure}

In this Letter, we report the experimental demonstration of $1{\rightarrow{3}}$ QTC and ODT of logical states using a four-qubit symmetric Dicke state 
with two excitations realized using a high-quality hyperentangled (HE) photonic resource~\cite{barbieri,dickeexpRome}. The entanglement-sharing structure 
of the state has been characterized quantitatively using 
a structural entanglement witness for symmetric Dicke states~\cite{structural,toth} and fidelity-based entanglement witnesses for the three- and two-qubit states achieved upon subjecting the Dicke 
register to proper single-qubit projections~\cite{dickeexp}. All such criteria have confirmed the theoretical expectations with a high degree of significance. As for the protocols 
themselves, the qubit state to teleclone/teleport 
is encoded in an extra degree of freedom of one of the physical information carriers entering such multipartite resource. This has been made possible by 
the use of a displaced Sagnac loop~\cite{alme07sci} [cf. Fig.~\ref{setup}], which introduced unprecedented flexibility in the setting, allowing for the realization of high-quality entangling two-qubit gates on 
heterogeneous degrees of freedom of a photon {\it within} the Sagnac loop itself. 
The high fidelities achieved between the experiments and theory 
(as large as $96\%$, on average, for ODT) 
demonstrate the usefulness of Dicke states as resources for distributed quantum communication beyond the 
limitations of a ``proof of principle". Our scheme is well suited for implementing $1{\rightarrow}N{>}3$ QTC of logical states or ODT with 
more than three receivers via the realization of larger HE resources, which is a realistic possibility. 

\noindent
{\it Resource production and state characterization.-} The building block of our experiment is the source of two-photon four-qubit polarization-path 
HE states developed in~\cite{barbieri,ceccarelli} and used recently to test multi-partite entanglement, decoherence and general quantum 
correlations~\cite{dickeexpRome,discord2011,chiur12njp}. Such apparatus has been modified as described in the Supplementary Information [SI]~\cite{epaps} to produce the HE state $\ket{\xi}_{abcd}{=}[\ket{HH}_{ab}(\ket{r \ell}-\ket{\ell r})_{cd}+2\ket{VV}_{ab}\ket{r \ell}]_{cd}/\sqrt6$. Here, we have used the encoding $\{\ket{H},\ket{V}\}{\equiv}\{\ket{0},\ket{1}\}$, 
with $H/V$ the horizontal/vertical polarization states of a single photon, and $\{\ket{r},\ket{\ell}\}{\equiv}\{\ket{0},\ket{1}\}$, 
where $r$ and $\ell$ are the path followed by the photons emerging from the HE stage~\cite{epaps}. Qubits $a,c$ ($b,d$) are encoded in the polarization and momentum of photon A (B). State $\ket{\xi}$ is turned into a four-qubit 
two-excitation Dicke state $\ket{D^{(2)}_4}{=}(1/\sqrt6)\sum^6_{j=1}\ket{\Pi_j}$ (with $\ket{\Pi_j}$ the elements of the vector of states constructed by taking all the permutation of $0$'s and $1$'s 
in $\ket{0011}$) by means of unitaries arranged as specified in Ref.~\cite{dickeexpRome}  [cf. Fig.~\ref{setup} {\bf a)}]. In the basis of the physical information carriers, the state reads
$\ket{D^{(2)}_4}{=}[\ket{HH\ell\ell}+\ket{VVrr}+(\ket{VH}+\ket{HV})(\ket{r\ell}+\ket{\ell r})]/\sqrt6$.
The fidelity of the protocols depends on the quality of this state, as will be clarified soon. 
We have thus tested the closeness of the experimental state to $\ket{D^{(2)}_4}$ and characterized its entanglement-sharing structure.  

First, we have ascertained the genuine multipartite entangled nature of the state at hand by using tools designed to assess the properties of symmetric Dicke states~\cite{structural,toth,campbell}. 
We have considered the multipartite entanglement 
witness 
\begin{equation}
\label{witness}
{\cal W}_m=[24\openone+\hat J^2_x\hat S_x+\hat J^2_y\hat S_y+\hat J^2_z(31\openone-7\hat J^2_z)]/12,
\end{equation}
which is specific of $\ket{D^{(2)}_4}$~\cite{toth} and requires only three measurement settings. Here, $\hat S_{x,y,z}{=}(\hat J^2_{x,y,z}-\openone)/2$ with $\hat J_{x,y,z}{=}\sum_{i\in{\cal Q}}\hat\sigma^{x,y,z}_i/2$ 
collective spin operators, ${\hat\sigma}^j~(j{=}x,y,z)$ the $j$-Pauli 
matrix and ${\cal Q}=\{a,b,c,d\}$. 
The expectation value of ${\cal W}_m$ is positive on any bi-separable four-qubit state, thus negativity implies multipartite entanglement. 
 Its experimental implementation allows to provide a lower bound to the state fidelity with the ideal Dicke state as 
$F_{D^{(2)}_4}\ge(2-\langle{\cal W}_m\rangle)/3$. When calculated over the resource that we have created in the lab, we achieve 
${\cal W}_m{=}-0.341\pm0.015$, which leads to  $F_{D^{(2)}_4}\ge(78\pm0.5)\%$. The genuine multipartite entangled nature of our state is 
corroborated by another significant test: we consider the witness testing bi-separability on multipartite symmetric, permutation invariant states like our $\ket{D^{(2)}_4}$~\cite{dickeexp,campbell} 
\begin{equation}
{\cal W}_{cs}(\gamma)=b_{4}(\gamma)\openone-(\hat{J}^2_x+\hat{J}^2_y+\gamma\hat{J}^2_z)~~~(\gamma{\in}\mathbb{R}).
\end{equation}
Here $b_{4}(\gamma)$ is the maximum expectation value of the collective spin operator $\hat{J}^2_x{+}\hat{J}^2_y{+}\gamma\hat{J}^2_z$ over the class of 
bi-separable states of four qubits and can be calculated for any value of the parameter $\gamma$.~\cite{campbell}. 
Finding $\langle{\cal W}_{cs}(\gamma)\rangle{<}0$ for some $\gamma$ implies genuine multipartite entanglement. The direct evaluation shows that already for $\gamma=-0.12$
the witness is negative by more than one standard deviation and by more than fifteen for $\gamma=-2.5$ 
(cf. SI~\cite{epaps}).

These results,  although indicative of high quality of the resource produced, are not exhaustive and further evidence is needed. 
In order to provide an informed and experimentally not-demanding analysis on the state being generated,  we have decided to resort to indirect yet highly significant 
evidence on its properties. In particular, we have exploited the interesting entanglement structure that 
arises from $\ket{D^{(2)}_4}$ upon subjecting part of the qubit register to specific single-qubit projections. In fact, by projecting one of the 
qubits onto the logical $\ket{0}$ and $\ket{1}$ states, we maintain or lower the number of excitations in the resulting state without leaving 
the Dicke space, respectively. Indeed, we achieve $\ket{D^{(2)}_3}=(\ket{011}+\ket{101}+\ket{110})/\sqrt3$ when projecting onto $\ket{0}$, 
while $\ket{D^1_3}{=}(\ket{100}+\ket{010}+\ket{001})/\sqrt3$ is obtained when the projected qubit is found in $\ket{1}$. Needless to say, 
these are genuinely tripartite entangled states, as it can be ascertained by using the entanglement witness formalism. For this task 
we have used the fidelity-based witness~\cite{acin} 
${\cal W}_{D^{(k)}_3}=({2}/{3})\openone{-}\ket{D^{(k)}_3}\bra{D^{(k)}_3}$ $(k=1,2)$,
whose mean is positive for any separable and biseparable three-qubit state, is $-1/3$ when evaluated over $\ket{D^k_3}$ and whose optimal 
decomposition (cf. SI~\cite{epaps}) requires five local measurement settings~\cite{acin, gueh03ijtp}. We have implemented the witness 
for states obtained projecting qubit $d$ ({\it i.e.} momentum of photon B), achieving $\langle{\cal W}^{exp}_{D^{(1)}_3}\rangle{=}-0.21{\pm}0.01$ 
and $\langle{\cal W}^{exp}_{D^{(2)}_3}\rangle{=}-0.24{\pm}0.01$ (the apex indicates their experimental nature) corresponding to lower bounds 
for the fidelity with the desired state of $0.876{\pm}0.003$ and $0.908{\pm}0.003$, respectively.

Finally, by projecting two qubits onto elements of the computational basis, one can obtain elements of the Bell basis. 
Indeed, regardless of the projected pair of qubits, $\bra{ij}D^{(2)}_4\rangle{=}\ket{\psi^+}$ with $\{\ket{\psi^\pm}{=}(\ket{01}{\pm}\ket{10})/\sqrt2,~\ket{\phi^\pm}{=}(\ket{00}{\pm}\ket{11})/\sqrt2\}$ the Bell basis and 
$i{\neq}{j}{=}{0,1}$. We have verified the quality of the reduced experimental states achieved by projecting the Dicke state onto
$\ket{10}_{cd}$ and $\ket{01}_{cd}$ using two-qubit quantum state tomography (QST)~\cite{jame01pra} on the remaining two qubits. By finding fidelities ${>}91\%$ regardless of the projections operated, we can claim to have a very good Dicke resource, which puts us in the position 
to experimentally implement the quantum protocols. 

\noindent
{\it $1{\rightarrow}3$ QTC and ODT.-} Telecloning~\cite{mura99pra} is a communication primitive that merges teleportation and cloning to deliver approximate copies of a quantum state to remote nodes of a network.  
Differently, ODT~\cite{bourennane} enables the teleportation of a state to an arbitrary location of the network. Both require shared multipartite entanglement. A 
deterministic version of ODT makes use of GHZ entanglement~\cite{panODT}, while the optimal resources for QTC are symmetric states having the form of superpositions of Dicke states with $k$ excitations~\cite{bourennane2010,bour06prl,mura99pra,paternostro2010}. Continuous-variable QTC 
was demonstrated in~\cite{koik06prl}. Although a symmetric Dicke state is known to be useful for such protocols (ODT being reformulated probabilistically)~\cite{kies07prl}, no experimental demonstration has yet been reported: in Ref.~\cite{kies07prl}, only 
an estimate of the efficiency of generation of a two-qubit Bell state between sender and receiver was given, based on data for $\ket{D^{(2)}_4}$. 
Differently, our setup allows to perform both QTC and probabilistic ODT. 


We start discussing the $1{\rightarrow}3$ QTC scheme based on $\ket{D^{(2)}_4}$, which is a 
variation of the protocol given in Ref.~\cite{mura99pra}. We consider the qubit state 
to clone $\ket{\alpha}_X=\alpha\ket{0}_X+\beta\ket{1}_X~(|\alpha|^2+|\beta|^2{=}1)$, held by a {\it client} $X$. 
The agents of a {\it server} composed of qubits $\{a,b,c,d\}$ and sharing the Dicke resource agree on the identification of a {\it port} qubit $p$.
The state of pair $(X,p)$ undergoes a Bell measurement (BM) performed by implementing a controlled-NOT gate ${\sf CX}_{Xp}$ 
followed by a projection of $X$ ($b$) on the eigenstates of $\hat\sigma^x$ ($\hat\sigma^z$).
They publicly announce the results of their measurement, which leaves us with 
$\bigotimes_{j{\in}{\cal S}_{tc}}{\sf P}_j(\alpha\ket{D^{(1)}_3}+\beta\ket{D^{(2)}_3})_{{\cal S}_{tc}}\otimes\ket{\psi_+}_{Xp}$,
where ${\cal S}_{tc}{=}\{a,b,c,d\}\slash p$ is the set of server's qubits minus $p$, $\ket{D^{(k)}_{3}}$ 
is a three-qubit Dicke state with $k{=}1,2$ excitations and the gates ${\sf P}_j$ (identical for all the qubits in ${\cal S}_{tc}$) are determined by the outcome of the BM, as illustrated in Fig.~\ref{setup} {\bf d)}. The protocol is now completed and the client's qubit is cloned into the state of the elements of ${\cal S}_{tc}$. To see this, we 
trace out two of the elements of such set and evaluate the state fidelity 
between the density matrix $\rho_{r}$ of the remaining qubit $r$ and the client's state, which reads 
${\cal F}(\theta){=}[9{-}\cos(2\theta)]/12$, where {$\alpha{=}\cos(\theta/2)$}. Clearly, the fidelity depends on the state to clone, achieving a maximum (minimum) of $5/6$ (2/3) at $\theta=\pi/2$ ($\theta=0,\pi$). This exceeds the value $7/9$ achieved by a universal symmetric $1\to3$ cloner due to the state-dependent nature of our protocol. 


\begin{figure}
\includegraphics[width=9cm]{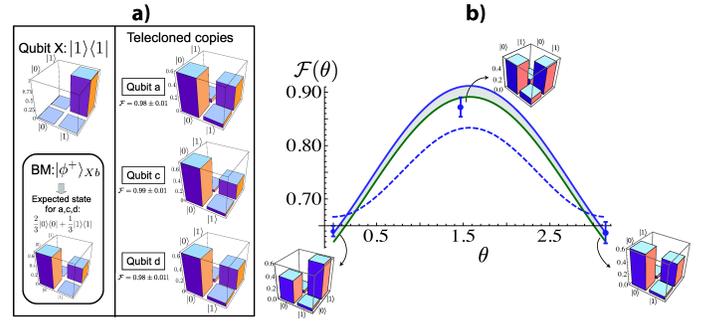}
\caption{(Color online) {\bf a)} Experimental QTC: for an input state $\ket{1}_X$, after the BM on $(X,b)$ with outcome $\ket{\phi^+}_{Xb}$, 
the ideal output state is 
$(2\ketbra{0}{j}{0}+\ketbra{1}{j}{1})/3$, $\forall j{=}a,c,d$ [left column of the panel]. 
The state of qubit $j$ after the experimental QTC, has very large overlap with the theoretical state. 
The right column of the panel shows the experimental single-qubit density matrices. {\bf b)} 
Theoretical QTC fidelity and experimental
density matrices of the clone (qubit $a$) for various input states. 
We show the fidelities between the experimental input states and clones (associated uncertainties determined by considering Poissonian fluctuations of the coincidence counts). The dashed line shows the theoretical fidelity for pure input states of the client's qubit. The dashed area encloses the values of the fidelity achieved for a mixed input state of $X$ and the use of an imperfect Dicke resource compatible with the states generated in our experiment (cf. SI~\cite{epaps})}
\label{Teleclone}
\end{figure}

We now introduce the ODT protocol. As for QTC, this is formulated as a game with a client and a server. The client holds qubit $X$, into which the state $\ket{\alpha}_X$ to teleport is encoded. The elements 
of the server share the $\ket{D^{(2)}_4}$ resource. The client decides which party $r$ of the server should receive the qubit to teleport 
($r$ and $p$ can be any of $\{a,b,c,d\}$, and  $r$ is chosen at the last step of the scheme). Unlike QTC, the client performs a ${\sf CX}_{Xp}$. At this stage the information on the qubit to teleport is {spread} across the server, and the client declares who will  receive it. Depending on his choice, the members in ${\cal S}_{odt}{=}\{a,b,c,d\}\slash\{r,p\}$ project their qubits onto $\ket{01}_{{\cal S}_{odt}}$, getting
$[\alpha(\ket{001}+\ket{010})_{Xpr}+\beta(\ket{111}+\ket{100})_{Xpr}]\otimes\ket{01}_{{\cal S}_{odt}}$.
{The scheme is completed by a projection onto $\ket{+1}_{Xp}$ with $\ket{+}{=}(\ket{0}+\ket{1})/\sqrt2$}~\cite{note}. 

\noindent
{\it Experimental implementations of $1{\rightarrow}3$ QTC.-} The setup in Fig.~\ref{setup} {\bf b)} and {\bf c)},  which represents a significant improvement over the scheme used in~\cite{dickeexpRome}, allows for the implementation of both the protocols. The shown displaced Sagnac loop and the use of the lower photon B allow us to add the client's qubit to the computational register. This is encoded in the sense of circulation of the loop by such field: modes $\ket{r}$ and $\ket{\ell}$ of photon B impinge on different points of beam splitter ${\rm BS}_2$, so that the photon entering the Sagnac loop can follow the clockwise path, thus being in 
the $\ket{{\circlearrowright}}{\equiv}\ket{0}$ state, or the counterclockwise one, being in $\ket{{\circlearrowleft}}{\equiv}\ket{1}$ (photon A does not pass 
through ${\rm BS}_2$). The probability $|\alpha|^2$ of being in 
the former (latter) state relates to the transmittivity (reflectivity) of ${\rm BS}_2$. {This probability is varied using intensity attenuators intercepting the output modes of ${\rm BS}_2$}. At this stage, the state of the register is  
$\ket{D^{(2)}_4}_{abcd}\otimes (\alpha \ket{\clock} + e^{i \phi_x}\sqrt{1- |\alpha|^2} \ket{\cclock})_X$,
where $\phi_x$ is changed by tilting the glass-plate in the loop. The ${\sf CX}_{Xp}$ gate has been implemented with qubit $X$ as the control, qubit $b$ ({\it i.e.} the polarization of photon B) as the port $p$ and taking a HWP rotated at $45^{\circ}$ with respect to the optical axes, placed only on the counterclockwise circulating modes of the Sagnac loop~\cite{notedelay}. The second passage of the lower photon in ${\rm BS}_2$ allows to project qubit $X$ on the eigenstates of $\hat\sigma^x_X$. To complete the Bell measurement on qubits $(X,p)$
we have placed a HWP and a PBS before the detector in order to project qubit $p$ on the eigenstates of $\hat\sigma^z_p$.
The remaining qubits ($a,c$ and $d$)
 embody three copies of the qubit $X$. Their quality has been tested by performing 
QST over the reduced states obtained by tracing over any two qubits. 
Pauli operators in the path DOF have been measured using the second passage of both photons through ${\rm BS}_1$.
The glass plates $\phi_{A,B}$ allowed projections onto $\frac{1}{\sqrt{2}}(\ket{r}+e^{i\phi_{A(B)}}\ket{\ell})_{c(d)}$. 
To perform QST on the polarization DOF we used an analyzer composed of HWP, QWP and PBS before the photo-detector. 
To trace over polarization, we removed the analyzer. To trace over the path, a delayer was placed on either $\ket{r}$ or $\ket{\ell}$ coming back to ${\rm BS}_1$, thus making them distinguishable and spoiling their interference.

In Fig.~\ref{Teleclone} {\bf a)} we show the experimental results obtained for the input states $\ket{1}_{X}$, when $p{=}b$. QST on qubit $j{=}a,c,d$
shows an almost ideal fidelity with the theoretical state, uniformly with respect to label $j$, thus proving the symmetry of QTC. Our setup allows us to teleclone arbitrary input states. {To illustrate the working principles and efficiency of the telecloning 
machine, we have considered the logical states $\ket{0}_X$ and $\ket{+}_X$ and $\ket{1}_X$ (i.e. we took $\theta{\simeq}0,\pi/2$ and $\pi$) and measured the 
corresponding copies in qubit $a$ ({\it i.e.} the polarization of photon A). States  $\ket{0}_X$ and $\ket{1}_X$ were generated by selecting the modes in the displaced Sagnac. In the first (second) case we considered only modes $\ket{\clock}$ ($\ket{\cclock}$), while $\ket{+}_X$ was generated using both modes and adjusting the relative phase with the glass-plate $\phi_X$ (by varying this phase, we can explore the whole phase-covariant case).   
Although the experimental results are very close to the expectations for ${\cal F}(\theta)$ [cf. Fig.~\ref{Teleclone} {\bf b)}], some discrepancies are 
found for $\theta=\pi/2$. In particular, the theory seems to underestimate (overestimate) the experimental fidelity of telecloning close to 
$\theta=\pi/2$ ($\theta=0,\pi$). These effects are due to the mixedness of the $X$ state entering the Sagnac loop as well as the suboptimal fidelity between the experimental resource and $\ket{D^{(2)}_4}$. 
In fact, the experimental input state corresponding to $\theta\simeq\pi/2$ has fidelity $0.91\pm0.02$ 
with the desired $\ket{+}_X$ due to depleted off-diagonal elements in its density matrix (cf. SI~\cite{epaps}). We have thus modelled the telecloning of dephased client states based on the use of a mixed Dicke channel of sub-unit fidelity with $\ket{D^{(2)}_4}$. 
The details are presented in Ref.~\cite{epaps}. Here we mention that, by including the uncertainty associated with the estimated $F_{D^{(2)}_4}$, we have determined a $\theta$-dependent region of telecloning fidelities into which the fidelity between the experimental state of the clones and the input client state falls. As shown in Fig.~\ref{Teleclone} {\bf b)}, this provides a better agreement between theory and data.
}

\noindent
{\it Experimental implementations of ODT.-} In ODT the client holds qubit $X$, which is added to the computational register using the Sagnac loop. The client's qubit has been teleported to the server's elements $a$ and $b$ ({\it i.e.} the polarization of photons A and B). The necessary ${\sf CX}_{Xp}$ gate has been implemented, as above, by taking $X$ as the control and $p{=}b$ as the target qubit. The server's elements $\{c,d\}$ have been projected onto $\ket{01}_{cd}$ and $\ket{10}_{cd}$.  Depending on the chosen receiver (either $a$ or $b$), the scheme is implemented by projecting onto $\ket{+1}_{Xa(b)}$ and performing QST of the teleported qubit $b(a)$. While the projection onto $\ket{+}_{X}$ has been 
realized using the second passage of the lower photon through ${\rm BS}_2$, a projection onto $\ket{1}_{a(b)}$ is achieved projecting the physical qubit onto $\ket{V}_{a(b)}$. In Table~\ref{ODT} we report the experimental results obtained for several measurement configurations and teleportation channels. 
In SI~\cite{epaps} we provide the reconstructed density matrices of qubits $\{X,a,b\}$ for each configuration used.

\begin{table}[t!]
\centering
\caption{Experimental fidelities between the
teleported qubit ($a$ or $b$) and the state of qubit $X$
(determined by $\theta$). Uncertainties result from associating
Poissonian fluctuations to the coincidence counts.}
\label{ODT}
\begin{tabular}{ccc||ccc}
\hline
Projection & $\theta$ & Fidelity & Projection & $\theta$ & Fidelity\\
\hline
$_{cd}\bra{10}$ & $0$    & $\mathcal{F}_{a}{=}0.93{\pm}0.01$ & $_{cd}\bra{01}$ & $\pi$  & $\mathcal{F}_{a}{=}0.98{\pm}0.01$\\
$_{cd}\bra{10}$ & $0$    & $\mathcal{F}_{b}{=}0.95{\pm}0.01$ & $_{cd}\bra{01}$ & $\pi$  & $\mathcal{F}_{b}{=}0.97{\pm}0.01$\\
$_{cd}\bra{01}$ & $0$    & $\mathcal{F}_{a}{=}0.97{\pm}0.01$ & $_{cd}\bra{10}$ & $1.46$ & $\mathcal{F}_{a}{=}0.92{\pm}0.02$\\
$_{cd}\bra{01}$ & $0$    & $\mathcal{F}_{b}{=}0.97{\pm}0.01$ & $_{cd}\bra{10}$ & $1.46$ & $\mathcal{F}_{b}{=}0.98{\pm}0.01$\\
$_{cd}\bra{10}$ & $\pi$  & $\mathcal{F}_{a}{=}0.96{\pm}0.01$ & $_{cd}\bra{01}$ & $1.37$ & $\mathcal{F}_{a}{=}0.97{\pm}0.02$\\
$_{cd}\bra{10}$ & $\pi$  & $\mathcal{F}_{b}{=}0.98{\pm}0.01$ & $_{cd}\bra{01}$ & $1.37$ & $\mathcal{F}_{b}{=}0.96{\pm}0.02$\\
\hline
\end{tabular}
\end{table}

    
\noindent
{\it Conclusions and outlook.-} We have implemented QTC and ODT of logical states using a four-qubit symmetric Dicke state. We have realized a novel setup based on the well-tested HE polarization-path states and complemented by  a displaced Sagnac loop. This allowed the encoding of non-trivial input states in the computational register, and the performance of high-quality quantum gates and protocols. Our results go beyond  state-of-the-art in the manipulation of experimental Dicke states and the realization of quantum networking.

 \begin{acknowledgments}
 {\it Acknowledgments.--} {We thank Valentina Rosati for the contribution given to the early stages of this work.
This work was supported by EU-Project CHISTERA-QUASAR, PRIN 2009 and FIRB-Futuro in ricerca HYTEQ,
and the UK EPSRC (EP/G004579/1).}
 \end{acknowledgments}


\newpage

\section{Supplementary Information on: Experimental Quantum Networking Protocols via Four-Qubit Hyperentangled Dicke States}


In this supplementary Information we provide further details on both the theoretical and experimental results and analysis reported in the main Letter.

\section{Resource production and state characterization} 

Here we describe the source of hyperentanglement that has been used as the building block of our experiment. As remarked in the text of the Letter, we use the encodings $\{\ket{H},\ket{V}\}{\equiv}\{\ket{0},\ket{1}\}$, with $H/V$ the horizontal/vertical polarization states of a single photon, and $\{\ket{r},\ket{\ell}\}{\equiv}\{\ket{0},\ket{1}\}$, where $r$ and $\ell$ are the path followed by the photons emerging from the HE stage introduced and exploited in~\cite{barbieri,dickeexpRome,discord2011,chiur12njp}. 

\begin{figure}
\includegraphics[width=\linewidth]{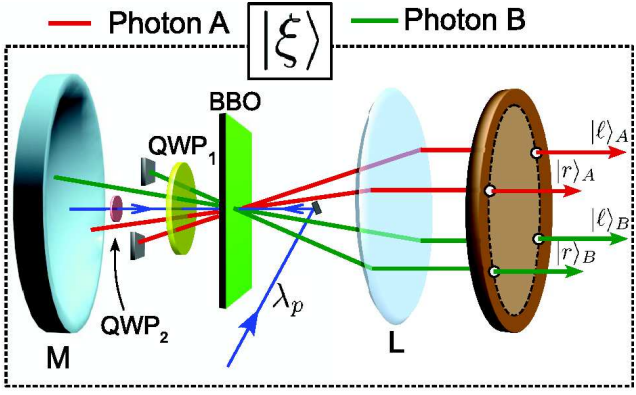}
\caption{Sketch of the experimental setup used to produce the HE resource state $\ket{\xi}_{abcd}$. The setup is discussed fully in the body of the text.}
\label{setupEPAPS}
\end{figure}

We modify such setup so to prepare the HE resource $\ket{\xi}_{abcd}{=}[\ket{HH}_{ab}(\ket{r \ell}-\ket{\ell r})_{cd}+2\ket{VV}_{ab}\ket{r \ell}]_{cd}/\sqrt6$ introduced in the main Letter. A sketch of the apparatus is shown in Fig.~\ref{setupEPAPS}. A Type-I nonlinear $\beta$-barium borate crystal, pumped by a vertically polarized laser field (wavelength $\lambda_p$), 
generates a polarization-entangled state given by the superposition of the spontaneous parametric down conversion (SPDC) signals at degenerate wavelength produced by a double-pass scheme. 
The mask selects four spatial modes $\{\ket{r},\ket{\ell}\}_{A,B}$ (two for each photon), parallelized by lens ${\rm L}$. ${\rm QWP}_{1,2}$ are quarter-wave plates. 
The first pass produces $2\ket{VV}\ket{r\ell}$. The spatial modes are intercepted by two beam stoppers. ${\rm QWP}_{1}$ changes the polarization into $\ket{VV}$ after 
reflection by mirror ${\rm M}$. The latter also reflects the pump, which produces the second-pass SPDC contribution $\ket{HH}(\ket{r\ell}-\ket{\ell r})$. The weight of this term 
in the final state  $\ket{\xi}$ is determined by ${\rm QWP}_{2}$~\cite{dickeexpRome}. 

\section{On entanglement witnesses for genuine multipartite entanglement}

Collective-spin operators are useful tools for the investigation of genuine multipartite entanglement, particularly for symmetric, permutation invariant states. One can construct the witness operator ~\cite{boundToth}
\begin{equation}
\label{standard}
{\cal W}_{cs}=b_{n}{\openone}-(\hat{J}^2_{x}+\hat{J}^2_{y}),
\end{equation}
where $b_{n}$ is the maximum expectation value of $\hat{J}^2_{x}+\hat{J}^2_{y}$ over the class of bi-separable states of $n$ qubits. Finding $\langle\hat{\cal W}^s_n\rangle\!<\!0$ for a given state implies genuine multipartite entanglement. It can be the case that Eq.~(\ref{standard}) fails to reveal the multipartite nature of a state endowed with a lower degree of symmetry. More flexibility can nevertheless be introduced by means of a suitable generalization such as 
\begin{equation}
\label{betterone}
\hat{\cal W}_{cs}(\gamma)=b_{n}(\gamma)\openone-(\hat{J}^2_{x}+\hat{J}^2_{y}+\gamma\hat J^2_z)~~(\gamma\in\mathbb R).
\end{equation}
Negativity of $\langle\hat{\cal W}_{cs}(\gamma)\rangle$ over a given state guarantees multipartite entanglement. The witness requires only three measurement settings and is thus experimentally very convenient. The bi-separability bound $b_{n}(\gamma)$ is now a function of parameter $\gamma$ and can be calculated numerically using the procedure described in Ref.~\cite{campbell}. In general, $b_{n}(\gamma)\!<\!b_{n}(0)$ for $\gamma<0$. Consequently, we restrict ourselves to the case of negative $\gamma$. 

In Table~\ref{tavola} we provide the experimental values of $\langle\hat J^2_{x,y,z}\rangle$ through which we have evaluated Eq.~\eqref{betterone}, which is plotted against $\gamma$ in Fig.~\ref{witness}. While $\langle\hat{\cal W}_{cs}(\gamma)\rangle$ soon becomes negative as $\gamma{<}-0.1$ is taken, the uncertainty associated with such expectation value, calculated by propagating errors in quadrature as
\begin{equation}
\label{error}
\delta \langle{\cal W}^{exp}_{cs}(\gamma)\rangle=\sqrt{\sum_{j=x,y}(\delta\langle{\hat J}^2_j\rangle^{})^2+\gamma^2(\delta\langle{\hat J}^2_z\rangle^{})^2},
\end{equation}
grows only very slowly with $\gamma$, therefore signaling an increasingly significant violation of bi-separability.

\begin{figure}
\includegraphics[width=6cm]{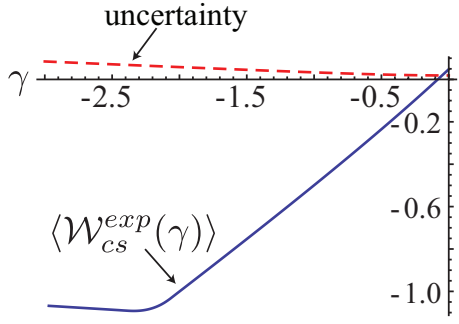}
\caption{Functional form of $\langle\hat{\cal W}_{cs}(\gamma)=\rangle$ against $\gamma$, as determined by the measured expectation values of collective spin operators (cf. Table~tavola). A negative value of $\langle\hat{\cal W}_{cs}(\gamma)\rangle$ signals genuine multipartite entanglement of the state experimental state under scrutiny. The associated experimental uncertainty [see Eq.~\eqref{error}] increases only very slowly as $|\gamma|$ grows. }
\label{witness}
\end{figure}

\begin{table}[b]
\caption{Experimentally measured expectation values of collective spin operators for the symmetric four-qubit Dicke state prepared in our experiment. The uncertainties are determined by associating Poissonian fluctuations to the coincidence counts.}
\centering
\begin{tabular}{c c }
\hline
\hline
Expectation value (with uncertainty) & Value \\
\hline
\hline
$\langle\hat J^2_x\rangle^{}\pm\delta\langle\hat J^2_x\rangle^{}$ & 2.568$\pm$0.015\\
$\langle\hat J^2_y\rangle^{}\pm\delta\langle\hat J^2_y\rangle^{}$ & 2.617$\pm$0.011\\
$\langle\hat J^2_z\rangle^{}\pm\delta\langle\hat J^2_z\rangle^{}$ & 0.039$\pm$0.028\\
\hline
\hline
\end{tabular}
\label{tavola}
\end{table}

\section{Optimal decomposition of the entanglement witness for $\ket{D^1_3}$}

As discussed in the main body of of the Letter, we have used a fidelity-based entanglement witness to characterize the genuine tripartite entanglement content of the state achieved upon projecting one of the qubits onto a state of the logical computational basis. Without affecting the generality of our discussion, here we concentrate on the case of a qubit-projection on qubit $d$ giving outcome $\ket{1}_d$, thus leaving us with state $\ket{D^{(1)}_3}_{abc}$. The fidelity-based witness that we have implemented is given in the main Letter and is decomposed in five measurement settings as~\cite{gueh03ijtp}
\begin{equation}
\begin{aligned}
{\cal W}_{D^{(1)}_3}&=\frac{1}{24}\Big\{17\openone{+}7\hat\sigma^z_a\hat\sigma^z_b\hat\sigma^z_c{+}3\hat\Pi[\hat{\sigma}^z_a\openone_{bc}]{+}5\hat\Pi[\hat\sigma_a\hat\sigma_b\openone_c]\\
&-\sum_{l=x,y}\sum_{k=\pm}(\openone_a{+}\hat\sigma^z_a{+}k\hat\sigma^l_a)(\openone_b{+}\hat\sigma^z_b{+}k\hat\sigma^l_b)(\openone_c{+}\hat\sigma^z_c{+}k\hat\sigma^l_c)\Big\}
\end{aligned}
\end{equation} 
where $\hat\Pi[\cdot]$ performs the permutation of the indices of its argument. The decomposition is optimal in the sense that ${\cal W}_{D^{(1)}_3}$ cannot be decomposed with lesser measurement settings. Experimentally, we have used the following rearrangement of the previous expression
\begin{equation}
\begin{aligned}
&{\cal W}_{D^{(1)}_3}{=}\frac{1}{24}\Big\{13\openone_{abc}{+}3\hat\sigma^z_a\hat\sigma^z_b\hat\sigma^z_c{-}\hat\Pi[\hat{\sigma}^z_a\openone_{bc}]{+}\hat\Pi[\hat\sigma^z_a\hat\sigma^z_b\openone_c]\\
&{-}2\hat\Pi[\hat\sigma^x_a\hat\sigma^x_b\openone_c]{-}2\hat\Pi[\hat\sigma^y_a\hat\sigma^y_b\openone_c]{-}2\hat\Pi[\hat\sigma^x_a\hat\sigma^x_b\hat\sigma^z_c]{-}2\hat\Pi[\hat\sigma^y_a\hat\sigma^y_b\hat\sigma^z_c]\Big\},
\end{aligned}
\end{equation} 
which was easier to implement with our setup. 

\begin{figure*}
\includegraphics[width=18cm]{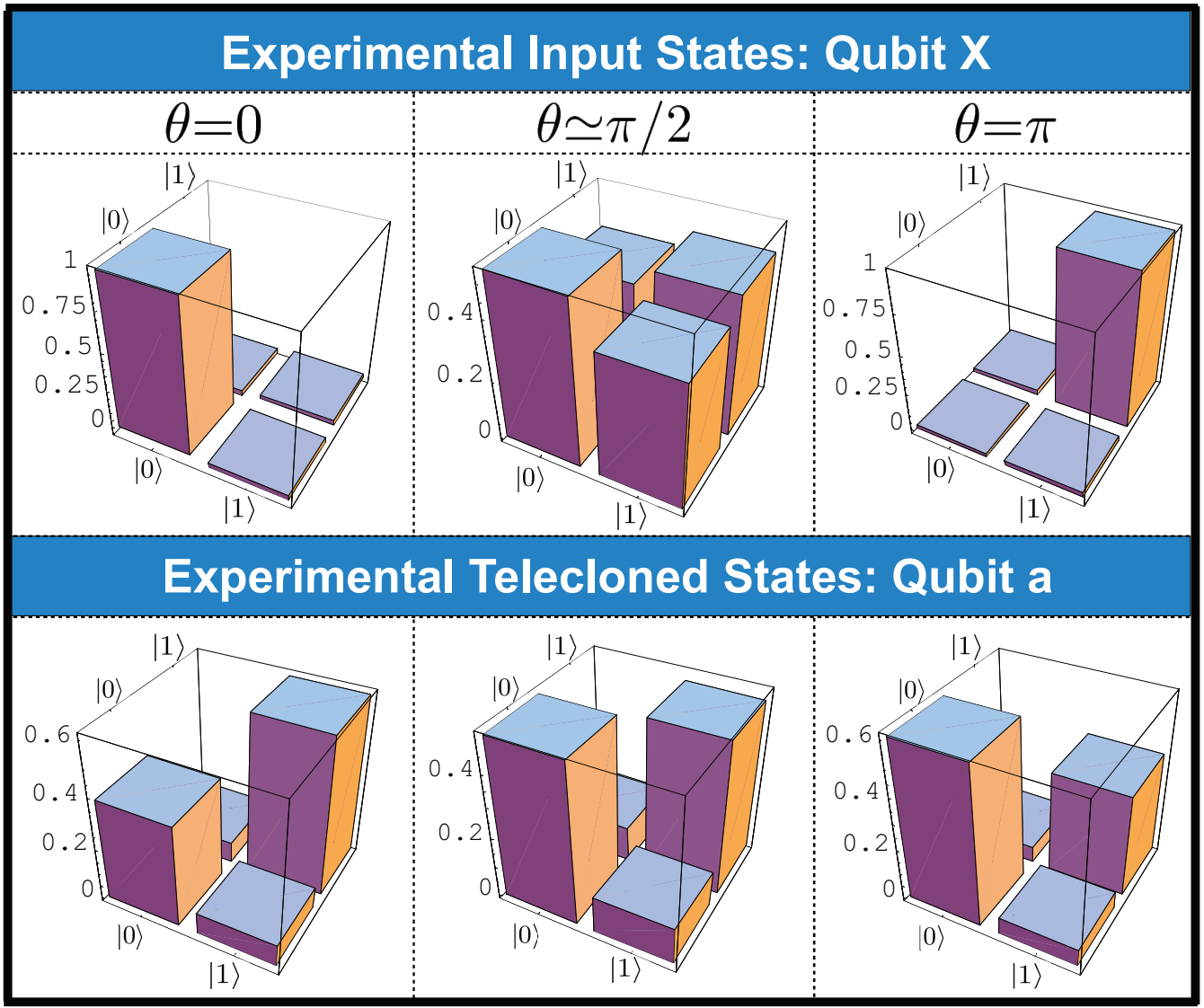}
\caption{We report the reconstructed density matrices of the telecloned states measured on qubit $a$, for three different input client's states (qubit $X$).}
\label{tomoQTC}
\end{figure*}

\begin{figure*}
\includegraphics[width=14cm]{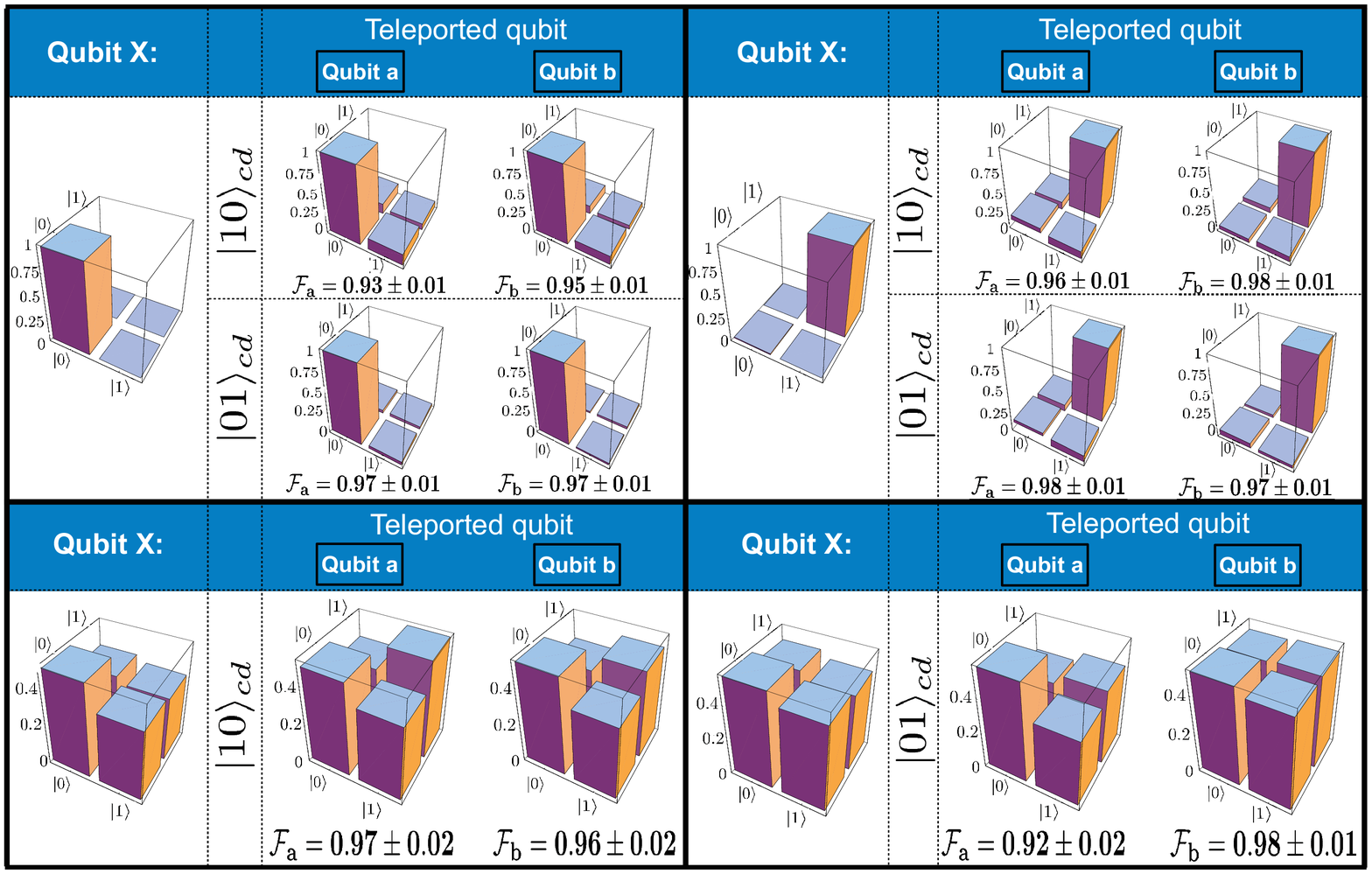}
\caption{We report the reconstructed density matrices of the various receiver states for four different input client's states (qubit $X$) and projections of the server's qubits onto both $\ket{01}_{{\cal S}_{odt}}$ and $\ket{10}_{{\cal S}_{odt}}$.}
\label{tomo}
\end{figure*}

\section{On the experimental measurement of the client's qubit for quantum telecloning}

A few remarks are in order on the way the client's qubit $X$ is experimentally measured in the 
actual implementation of the quantum telecloning protocol.

Due to slight unbalance at BS2 of Fig. 1 {\bf c)} of the main Letter, the blue and yellow paths in the Sagnac loop
used in order to encode qubit $X$ are unbalanced. We have thus corrected for such an asymmetry by first measuring 
the state of qubit $X$ generated entering the loop only with $\ket{r}$ modes [{\it i.e.} the blue path in Fig. 2 {\bf c)} and {\bf d)}].
We have then done the same with the $\ket{\ell}$ modes (yellow paths). Finally, we have traced out the path degree of freedom 
embodied by $\{\ket{r},\ket{\ell}\}$ by summing up the corresponding counts measured for every single projection that is needed 
for the implementation of single-qubit quantum state tomography, therefore reinstating symmetry. 

\section{Fidelity of quantum telecloning for mixed states of the client}

Here we provide a model for  the solid (red) line of Fig.~2 {\bf (b)} accounting for the fidelity of quantum telecloning of a client's mixed state. The evaluation of the theoretical fidelity of telecloning given in the main Letter does not take into account the mixed nature of the client's state, as well as the non-ideality of the experimental Dicke channel used for the scheme. As argued in the main Letter, these are the main sources of discrepancy between the experimental results and the theoretical predictions. Here we provide a simple model that includes these imperfections and allows for a more faithful comparison between theoretical predictions and experimental data. 

Our starting point is the observation that mixed input states of the client can correspond to telecloning fidelities larger than the theoretical values predicted by ${\cal F}(\theta)=[9-\cos(2\theta)]/12$. This can be straightforwardly seen by running the quantum telecloning protocol with a decohered state resulting from the application of a dephasing channel to a pure client's state of the form $\alpha\ket{0}_X+\beta\ket{1}_X$ with $\alpha=\cos(\theta/2)$ as in the main Letter. This is illustrated in Fig.~2 of the main Letter. Quite intuitively, as the input client's state loses tis coherences, the fidelity of telecloning improves. The second observation we make is that the entangled channel used in our experiment, although being of very good quality, has a non-unit overlap with an ideal Dicke resource. Taking into account the major sources of experimental imperfections, along the lines of the investigation in~\cite{dickeexpRome}, a reasonable description of the four-qubit resource produced in our experiment is the Werner-like state
\begin{equation}
\label{wernerlike}
\rho_{D}=p\ket{D^{(2)}_4}\bra{D^{(2)}_4}+(1-p){\openone}/16
\end{equation}
with $0\le p\le1$. The entangled Dicke component in such state is evaluated considering that our experimental estimate for the lower bound on the state fidelity is$F_{D^{(2)}_4}=(0.78\pm0.5)$. Moreover, we have checked that slight experimental imperfections in the determination of the populations of the input client's states (within the range observed experimentally) do not affect the overall picture significantly. We have thus incorporated the effects of a coherence-depleted input states of qubit $X$ into the protocol for $1\rightarrow3$ quantum telecloning performed using a mixed Dicke resource as in Eq.~(\ref{wernerlike}). The dephasing parameter used in the model for mixed client's state has been adjusted so that, at $\theta=\pi/2,$ we get the real part of the experimentally reconstructed off-diagonal elements of the density matrix of qubit $X$ (fixed relative phases between $\ket0_X$ and $\ket1_X$ do not modify our conclusions). The resulting state fidelity, shown in Fig. 2 {\bf b)} of the main letter, shows a very good agreement with the experimental data.

\section{Single-qubit quantum state tomography of receivers' states in experimental QTC and ODT}

In Fig.~\ref{tomoQTC} (Fig.~\ref{tomo}) we give the single-qubit density matrix obtained through quantum state tomography of the receiver's state in the QTC (ODT) protocol.
The telecloned states reported in Fig.~\ref{tomoQTC} have been shown in Fig.~2 {\bf b)} of the main letter. We have considered three different input client's 
states. For each of them, we have measured the telecloned state on the qubit $a$.     

The values of state fidelity included in the Figure~\ref{tomo} are those reported in Table I of the main Letter. We have considered four different input client's 
states. For each of them, we have projected the server's elements onto either $\ket{01}_{{\cal S}_{odt}}$ or $\ket{10}_{{\cal S}_{odt}}$ and taken 
qubit $a$ or $b$ as he receiver. The corresponding quantum state fidelities are evidently quite uniform and consistently above $90\%$ (mean fidelity $0.96\pm0.01$),
thus demonstrating high-quality and receiver-oblivious ODT.

\end{document}